*Brief communication*

# Problems with the use of Web search engines to find results in foreign languages

Dirk Lewandowski

Hamburg University of Applied Sciences, Faculty Design, Media and Information, Department Information, Berliner Tor 5, D – 20249 Hamburg, Germany

E-mail: dirk.lewandowski@haw-hamburg.de

**Abstract**

**Purpose** – To test the ability of major search engines, Google, Yahoo, MSN, and Ask, to distinguish between German and English-language documents
**Design/methodology/approach** – 50 queries, using words common in German and in English, were posed to the engines. The advanced search option of language restriction was used, once in German and once in English. The first 20 results per engine in each language were investigated.
**Findings** – While none of the search engines faces problems in providing results in the language of the interface that is used, both Google and MSN face problems when the results are restricted to a foreign language.
**Research limitations/implications** – Search engines were only tested in German and in English. We have only anecdotal evidence that the problems are the same with other languages.
**Practical implications** – Searchers should not use the language restriction in Google and MSN when searching for foreign-language documents. Instead, searchers should use Yahoo or Ask. If searching for foreign language documents in Google or MSN, the interface in the target language/country should be used.
**Value of paper** – Demonstrates a problem with search engines that has not been previously investigated
**Paper type** – Research paper

**Introduction and literature review**

Search engines offer many search options, usually through their advanced search interfaces (Lewandowski, 2004a; Notess, 2006). While these search options are not employed often by the majority of users (Jansen & Spink, 2006; Machill, Neuberger, Schweiger, & Wirth, 2004; Schmidt-Maenz & Koch, 2006), they often provide the only method for successfully conducting certain queries. Information professionals, for example, rely on advanced search options in order to conduct their research.

However, the problem with search features' operational reliability is often overlooked. While functions with few or no problems exist (such as restrictions to the top-level domain), other functions that are relatively easy to apply do not work properly in some major search engines (e.g., Boolean OR in Google, see Notess, 2000). Features such as language restrictions, searches for related pages, content filters, and date restrictions (see Lewandowski, 2004b), are more difficult to determine how well they work in different search engines.

The language of the target document is important because it determines whether the user can understand the search results. There are some issues with specific languages that search engines have to face (Guggenheim & Bar-Ilan, 2005; Lazarinis, 2007; Moukdad & Cui, 2005). However, for our investigation, we do not concentrate on these processing issues; instead, we focus on the language-identification aspect of search engines.

Search engines consider language factors when ranking results. Results in languages different from the language of the interface receive a lower ranking. While the result sets for a certain query are the same e.g. in the German and in the English versions of Google, the rankings are different. However, these rankings should not be confused with geographically dependent rankings, which employ the location of the user (as determined by his IP address or the country interface) and provide a higher ranking for documents from the user's country than from foreign countries. For example, this feature is used for Spanish-language documents from South America on the search engines' interfaces in Spain.

In order to help solve this problem, a user can choose the target language of the search engine's interface. For instance, a German user searching for English-language documents could use the .co.uk interface to receive a higher ranking for English-language documents. The use of this interface does not exclude all documents in other languages but ranks them below the documents in the target language of the interface. A properly functioning language restriction would exclude these documents. However, we identified cases in which search engines provided search results in a certain language, even though the language was explicitly excluded in the search statement. These instances presented surprising results because the automatic identification of a document's language from the text is a solved problem (Cavnar & Trenkle, 2004; Dunning, 1994; Grefenstette, 1995), although certain algorithms for specific characteristics of Web text must be adjusted (Martins & Silva, 2005). Besides the text, other factors can assist in the detection of a Web document's language, such as the location of the server and the surrounding documents.

The present study is in the context of our framework for measuring search engine quality (Lewandowski & Höchstötter, 2008), which specifies four areas of quality, including the quality of index, results, search functions, and the usability of search engines. However, this particular study focuses on only one aspect of the search functions. However, we believe that the language restriction plays an integral part in the overall measurement of search engine quality.

**Methods**

For our investigation, we collected 50 words that are used in German and in English as well. In choosing the words, we did not consider the word's definition; therefore, the same word might have different meanings in German and in English. For example, we used the word "handy," which means "mobile phone" in German but consists of different definitions in English.

We used the language-restriction feature on the advanced search pages of the following search engines: Google, Yahoo, MSN/Live[1], and Ask. We chose these engines because each of them offers its own index and international coverage. Details on selecting the major search engines can be found in Lewandowski (2008).

We performed each query twice with a restriction to either German or English and used the German-language interface in both cases. We checked the first 20 results for each query and marked each result as either German or English. In some cases, such as when the pages did not provide textual information, we were not able to detect the language; however, in other instances, both languages were included on the same page. Data was collected on December 8[th] and 9[th] of 2007.

**Results**

Table 1 illustrates an overview of the results. The data demonstrates that none of the engines faces significant problems in delivering German-language pages when using the German interface. However, in the case of an English-language restriction, the engines provide a variety of results. While Yahoo and Ask offer few results in languages other than English (approximately one result in a set of 20), Google and MSN encounter major problems in providing results in English. Google, for example, provides more than half of the results in the incorrect language (i.e., German), while MSN confronts this problem with more than two-thirds of the results.

Table 1: Percentage of one-language pages correctly assigned in the top 20 results

| Search engine | Target language: German | Target language: English |
|---|---|---|
| Ask | 99% | 94% |
| Google | 98% | 46% |
| MSN | 98% | 34% |
| Yahoo | 100% | 95% |

Upon analyzing these results, we hypothesize that Google and MSN do not use static-language detection, but instead use graded-language detection, in which a certain probability that a document contains a specific language is assigned to each document. This could mean that a document that (presumably) includes content in different languages is assigned to more than one language, but with a different percentage for each. As we know that documents in the target language of the interface used are boosted in the ranking, it is probable that documents that are at least in part written in the interface language are boosted in the ranking, even if they just contain some words that are assigned to the interface language. E.g., this could be the case with documents that contain a certain amount of words of foreign origin.

In our investigation, language could not be assigned to some documents because they either contained English and German text or only consisted of a language-selection option. In the case of Google, a total of 124 such pages existed in the English-language restriction, while only 30 pages existed in the German-language restriction. These results support our presumption that Google uses graded language detection.

For documents in which no language could be assigned, the situation with MSN was quite different. While 40 pages surfaced in the English selection, which included content in German and in English (or just a language-selection option), different MSN pages appeared without any text (but with pictures, sound, or video) or with a few words that could not accurately determine the language (such as names, phone number or address information). These results lead us to conclude that MSN either relies heavily on text-independent factors in determining the language or that a few sample pages determine the language of an entire Web site. Considering individual queries, we are not able to find patterns that could explain why in the individual cases, Google or MSN produce a certain amount of mismatches.

Fig. 1 provides a graph for the individual search engines. The x-axis includes the number of results that are considered, and the y-axis features the percentage of correctly assigned documents that use English as the target language. The graph supports the conclusion that Google heavily boosts the German-language pages: The more results positions are considered, the higher the ratio of correctly assigned documents. Looking at just the first results position, we recognize that Google presents an English-language result in only 19 percent of the cases. The same holds true for MSN, but to a much smaller degree.

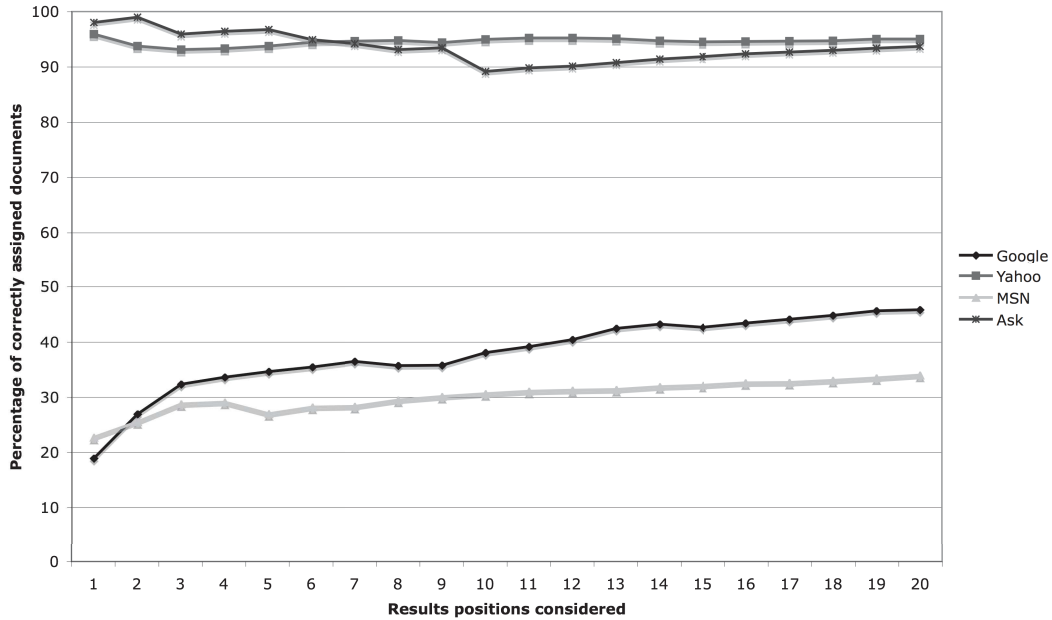

Fig. 1: Percentage of English pages according to results position

**Conclusion**

Our study indicates that search engine users should not use Google or MSN to conduct a language restriction in a foreign language when using an interface that is accustomed to a different language. Instead, a searcher should use the interface in the target language or the interface in the target country, if different countries use the same language. However, this strategy does not solve the problem of excluding results in foreign languages. While we were only able to prove the difficulties in using Google and MSN with the language pair German-English, pre-tests suggest that the problems are the same for other languages.

How can search engines provide language-restricted search functionality? From a technical standpoint, the detection of the correct language of a document is not a problem. Sufficient detection algorithms should be used, and documents to which more than one language is assigned should be carefully selected when the searcher uses a language restriction. Search engines should provide only results that match the language restriction 100 percent.

---

[1] MSN changed the name of its search engine to "Live.com". However, as the name of the Microsoft's search portal is still MSN and the users are used to this name, we stick to it.